\documentstyle[epsfig,longtable]{aipproc}

\begin{document}
\title{Time Resolved GRB Spectroscopy\footnote{Paper presented at the 5th 
Huntsville Symposium, Huntsville (Alabama), 19 - 22 October 1999.}} 
\author{Marco Tavani$^*$,
 David Band$^{\dagger}$,
 Giancarlo Ghirlanda$^*$ }
%  Marco Feroci$^{\ddagger}$}
\address{
$^*$Istituto Fisica Cosmica -- CNR,  Milan (Italy)\\
$^{\dagger}$X-2, Los Alamos National Laboratory, Los Alamos, NM 87545}

\maketitle

\begin{abstract}
We present the main results of a study of
time-resolved spectra of 43 intense GRBs detected by BATSE. 
We considered the 4-parameter Band model and the 
Optically Thin Synchrotron Shock model (OTSSM). 
We find that the large majority of time-resolved spectra of GRBs
are in remarkable agreement with the OTSSM. 
However, about  15 \% of {\it  initial  GRB pulses} 
show an apparent low-energy photon suppression. 
This phenomenon indicates that complex radiative
 conditions modifying optically thin emission 
 may occur during the initial phases of some GRBs.

\end{abstract}

\section*{Introduction}

We study a sample of 43 GRBs selected for the high-quality 
of their time-resolved  spectra obtained
with the  BATSE Spectroscopy Detectors 
(sensitive in the energy range $25 - 1800$~keV).
The time over which each spectrum
was accumulated was varied so that the signal-to-noise ratio was greater
than 15 (in the hard X-ray energy band).  
These data provide excellent temporal resolution: in many cases
we obtain more than 10 spectra per burst with accumulation times as short
as 256 ms.

\section*{Spectral Models}

We fitted each GRB time-resolved spectrum 
  with two  models: (1)  the Band model [1],  and 
(2) the Optically Thin Synchrotron Shock Model (OTSSM) 
\cite{A:mtav:3,A:mtav:4}.
The (purely phenomenological) 
4-parameter Band model \cite{A:mtav:1} 
consists of  two power-law components 
(of spectral indexes $\alpha$ and $\beta$) 
joined smoothly by an exponential roll-over near a break 
energy $E_{0}$.

\begin{eqnarray}
N(E) & = & A {\left( \frac{E}{{\rm 100 \, keV}} 
\right)^\alpha} \exp \left( - \frac{E}{E_{0}} \right) 
\;\;\;\;\;\;\;\;\;\;\;\;\;\;\;
\;\;\;\;\;\;\;\;\;\;\;\;\;\;\;\; \; \; \; {\rm for} \;\;
 E \leq \left( \alpha - \beta \right) E_{0}   \\
% \nonumber \\
N(E) & = & \left[A {\left( \frac{ \left( \alpha - \beta \right) 
E_{0}}{{\rm 100 \, keV}} \right)^{\alpha - \beta}} 
\exp \left( \beta - \alpha \right)\right] \left(\frac{E}{{\rm 100 \, keV}} 
\right)^\beta 
 \; \; \; {\rm for} \;\;
 E \geq \left( \alpha - \beta \right) E_{0}  
\end{eqnarray}

We used the (three-parameter)  OTSSM of Refs.
\cite{A:mtav:3,A:mtav:4}. 
We performed an independent spectral fitting for the Band 
and OTSSM models for each of the time-resolved spectra of all 
GRBs of our sample.
For each GRB we obtain 4 (3) best fit parameters
as a function of time
 %(actually it is the integration time of every time-resolved spectrum) 
representing  the complete  spectral evolution.

\section*{Results}

We find GRB spectral evolutions of two types:
(1)   a ``tracking behaviour", with spectral parameters in approximate
one-to-one correspondence with the changing energy flux,
and (2) a ``hard-to-soft evolution", with  spectral parameters evolving 
independently of the energy flux (see, e.g., ref. \cite{A:mtav:2}).
 
\begin{figure}[b!] % fig 1
\centerline{\epsfig{file=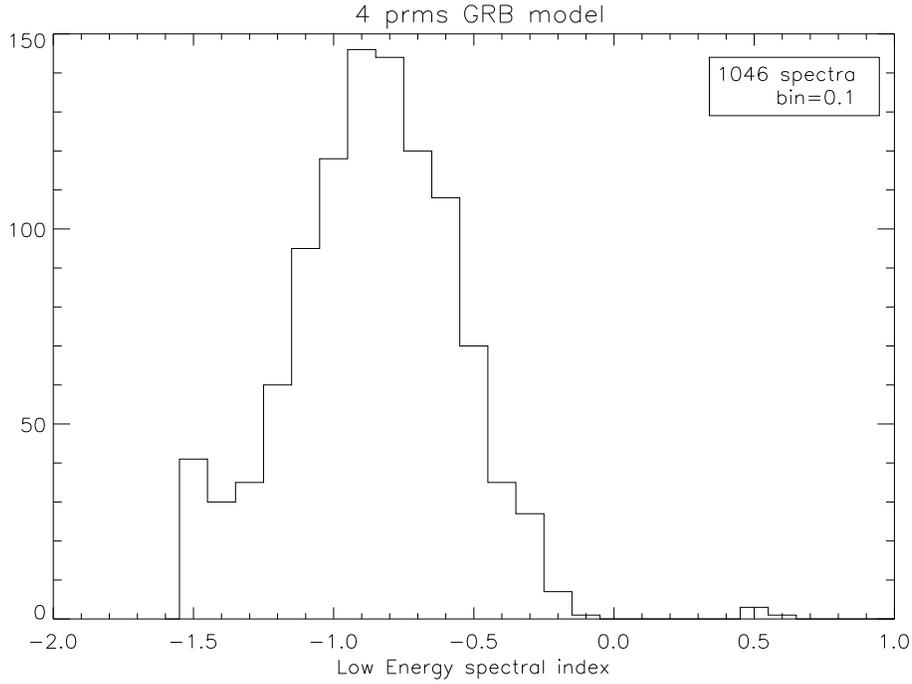,height=9cm,width=12cm}}
%\vspace*{9cm}
%\vspace{0.01pt}
\caption{Low-energy
($\alpha$) spectral index distribution from all the time-resolved spectra}
\label{GP-20-fig01}
\end{figure}

\begin{figure}[b!] % fig 2
\centerline{\epsfig{file=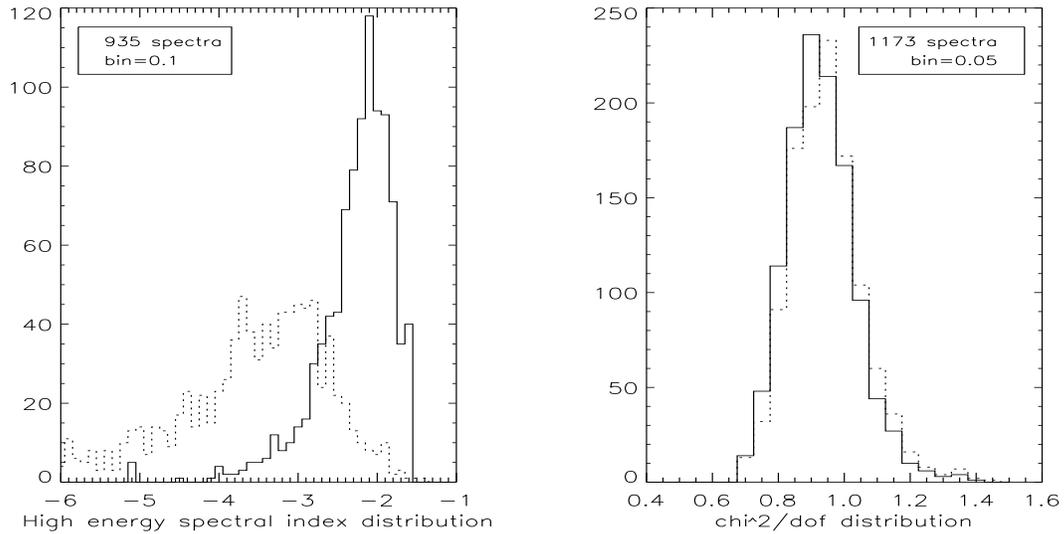,height=7cm,width=14cm}}
%\vspace*{7cm}
%\vspace{1pt}
\caption{\textit{Left panel:} High energy spectral index 
$\beta$ distributions for the Band model (\textit{solid line})
 and the OTSSM (\textit{dotted line}). 
\textit{Right panel:} Reduced $\chi^2$ distributions.}
\label{GP-20-fig02}
\end{figure}

Fig.\ref{GP-20-fig01} shows the distribution for \textit{all} 
collected time-resolved spectra of the 
%3-$\sigma$ values of the 
low-energy spectral index $\alpha$. A 
few bursts show values  $\alpha \ge$ --2/3 typically during the 
initial-rising part of their most intense pulses.
The high energy spectral index $\beta$ is less constrained, 
and in some cases varies substantially over consecutive spectra within 
the same burst. The $\beta$ distribution (Fig.\ref{GP-20-fig02} -- left
panel) 
is peaked near --2 for the Band model representation, and is
broader for the OTSSM fits.
Break energies $E_{0}$ are typically well below 500 keV. 
Interestingly, we find that the OTSSM provides a very good representation
of time-resolved spectral data. 
Fig.\ref{GP-20-fig02} (right panel) show the cumulative distribution
 of the reduced $\chi^2$ for the Band and OTSSM models.

\section*{Discussion}

We studied 43 GRBs from the BATSE spectral archive 
selected by their large signal-to-noise ratios. 
We collected information for a total of 1046 spectra. 

\begin{figure}[b!] % fig 4
\centerline{\epsfig{file=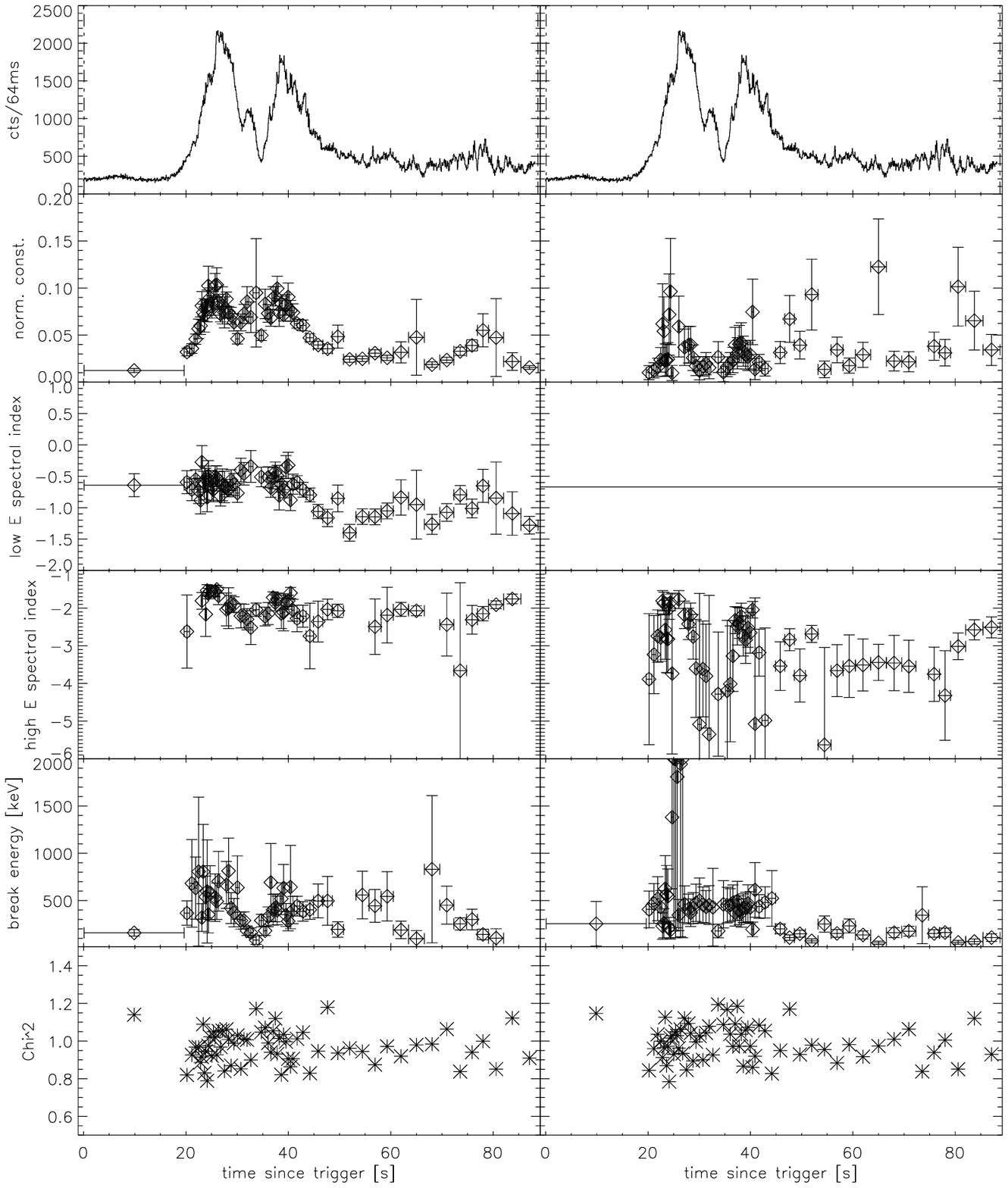,height=18cm, width=14cm}}
%\vspace*{18cm}
%\vspace{2pt}
\caption{GRB~990123 spectral evolution of the 4-parameter 
Band model (\textit{left column}) and the 3-parameter
 OTSSM (\textit{right column}). The $\alpha$ parameter is 
fixed in the OTSSM.}
\label{GP-20-fig03}
\end{figure}

\begin{figure}[b!] % fig 3
\centerline{\epsfig{file=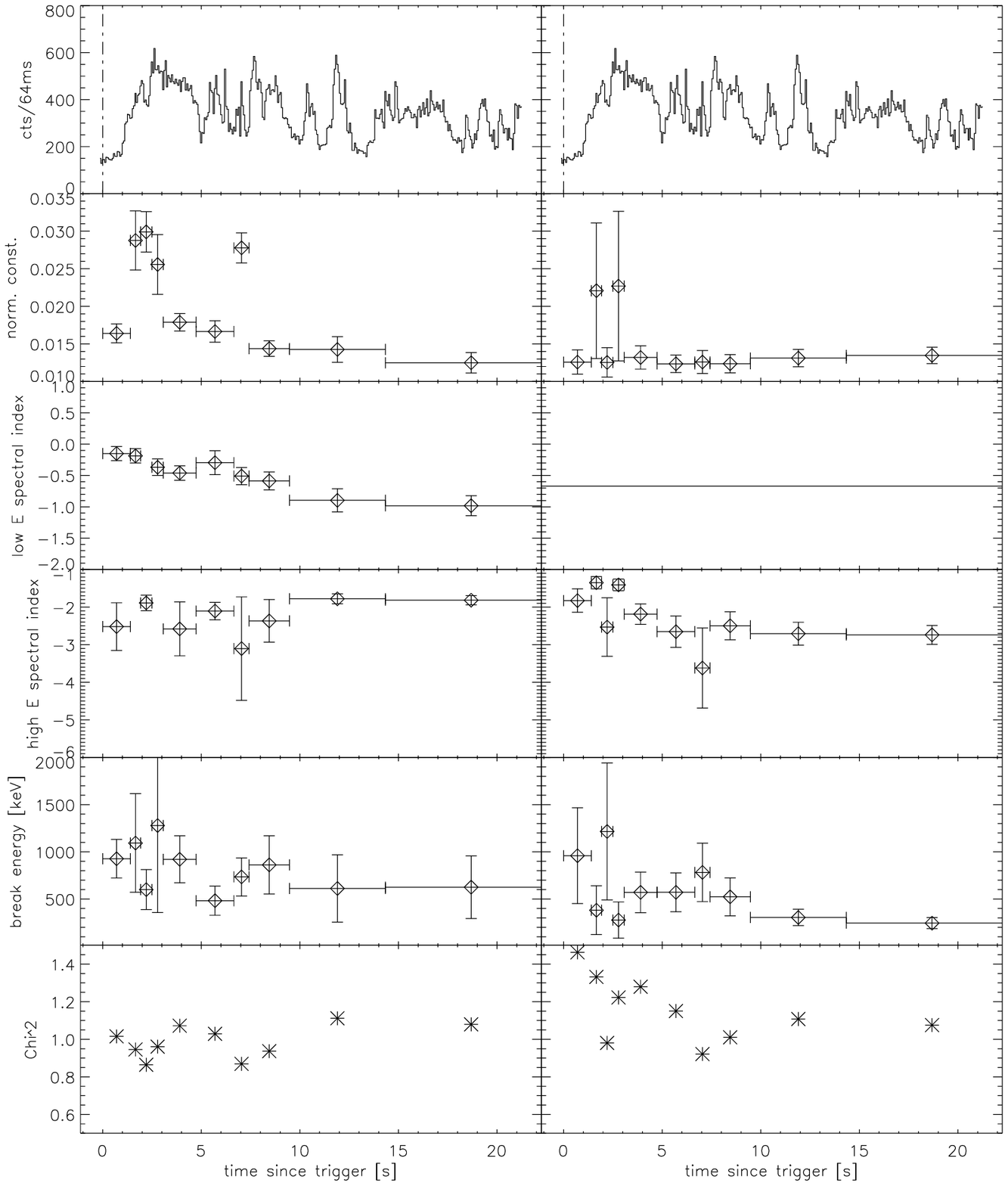,height=18cm, width=14cm}}
%\vspace*{18cm}
%\vspace{2pt}
\caption{GRB~910814 spectral evolution 
of the 4-parameter Band model (\textit{left column}) and the 3-parameter
 OTSSM (\textit{right column}). The $\alpha$ parameter is fixed in the
OTSSM.}
\label{GP-20-fig04}
\end{figure}

Our results indicate that the OTSSM is quite successful 
in describing the majority of GRB spectra.
Fig.\ref{GP-20-fig03} shows the spectral evolution of the remarkable
GRB~990123 demonstrating the validity of the  OTSSM for
 very intense bursts. 
However, violations of the simple OTSSM are apparent in about 
$15\%$ (at $3\sigma$ level) of our time resolved spectra. 
These violations (typically with a low-energy index $\alpha > -2/3$) 
always occur at the beginning of major GRB pulses (as in
Fig.\ref{GP-20-fig04}). 

The OTSSM was derived \cite{A:mtav:3} for idealized plasma 
and hydrodynamic conditions that are most likely
valid far from the central source. 
Several plasma and dynamic conditions (probably 
involving emission sites close to a central object) may produce 
the apparent suppression of soft photons at the beginning of
some GRB pulses.

\end{document}